\documentclass[
  ,final            
  ]
  {aipproc}

\layoutstyle{6x9}

\pdfoutput=1

\usepackage{amsmath}
\usepackage{amssymb}
\usepackage{dsfont}
\usepackage{color}
\newcommand{\bC}{\mathds C}
\newcommand{\bE}{\mathds E}
\newcommand{\bN}{\mathds N}
\newcommand{\bR}{\mathds R}
\newcommand{\cA}{\mathcal A}
\newcommand{\cB}{\mathcal B}
\newcommand{\cC}{\mathds M}
\newcommand{\cE}{\mathcal E}
\newcommand{\cS}{\mathcal S}
\newcommand{\id}{\mathds{1}}
\newcommand{\ii}{\rm i}
\newcommand{\sa}{\cA_{\rm sa}}
\begin{document}

\title{Discontinuities in the Maximum-Entropy Inference}


\author{Stephan Weis}{address={
Max Planck Institute for Mathematics in the Sciences,\\
Inselstrasse 22, D-04103 Leipzig, Germany}
}



\begin{abstract}
We revisit the maximum-entropy inference of the state of a finite-level quantum system
under linear constraints. The constraints are specified by the expected values of a set 
of fixed observables. We point out the existence of discontinuities in this inference 
method. This is a pure quantum phenomenon since the maximum-entropy inference is 
continuous for mutually commuting observables. The question arises why some sets of 
observables are distinguished by a discontinuity in an inference method which is 
still discussed as a universal inference method. In this paper we make an example of
a discontinuity and we explain a characterization of the discontinuities in terms of 
the openness of the (restricted) linear map that assigns expected values to states.
\end{abstract}

\maketitle


\section{Introduction}
\par
Methods of maximizing entropies have a long tradition. Typically, partial information 
is available, and a unique inference is desired, in terms of a probability distribution 
describing the state of a system under observation. The universality of entropy methods 
is a controversial topic, but from the perspective of the present article, it makes the 
question interesting why some sets of observables are distinguished through the 
discontinuity of an entropic inference method. Another important question is whether 
such observables can be implemented in physical experiments. We have no answers to both 
questions. In this article we describe a condition of continuity which will certainly 
be helpful to find more examples theoretically. We stress that this discussion takes 
place in a finite-dimensional matrix algebra, where entropy functionals are continuous. 
The Shannon and von Neumann entropies will not be considered in infinite settings 
\cite{Ho09,Shirokov10}.
\par
Let us dwell upon the idea of universality. The maximization of the Shannon entropy 
under linear constraints dates back to 1877 with Boltzmann's work about the energy 
distribution of a particle in a gas. See for example
\cite{Uffink06,Giffin08,Lesne11,Caticha12}. Jaynes \cite{Jaynes1} argued that the 
axioms of the Shannon entropy prove that the method of maximizing entropy is the least 
biased inference possible if partial information is available. Shore and Johnson 
\cite{Shore-Johnson} moved on to axiomatize not the entropy but the inference under
partial information, leading to the method, later termed {\it ME}, of minimizing the 
Kullback-Leibler divergence under the constraint of partial information, relative to 
a prior probability distribution describing the state of the system. 
Later on Skilling \cite{Skilling} argued that ME is a universal method of updating 
a probability distribution (or a positive distribution) given new information in terms 
of a constraint. The universality comes from the idea of {\it induction} in the 
philosophical sense, that is, finding a theory or a general rule from examples. See also 
\cite{Caticha04,Giffin08,Caticha12} and see Csisz\'ar \cite{Csiszar91} for axioms of
ME under linear constraints, where continuity is postulated (for distributions of full
support). Critics of the universality of ME claim that a (relative) {\it Renyi-entropy} 
should be minimized in inference \cite{Karbelkar86,Uffink95} and that the entropy 
function depends on a choice of properties that should be preserved during updating the 
prior \cite{Caticha04}. A counterattack against the Renyi-entropies is \cite{Caticha06}.
The universality of ME is doubted in particular outside of physics where the wide range 
of mathematical results about the Shannon entropy, from coding theorems over
data compression and much more, {\it ``provide a rationale, as well as several caveats, 
to the maximum entropy principle''} \cite{Lesne11}. The Kullback-Leibler divergence
generalizes some mathematical properties of the Shannon entropy, like a game theoretical 
aspect \cite{Topsoe79,Gruenwald04} or a data compression property \cite{Benedetto10}. 
After all, maximum-entropy methods have a wide range of applications, as documented 
for example in the conference proceedings of {\it Maximum Entropy and Bayesian Methods},
which make their analysis important.
\par
Von Neumann \cite{vonNeumann27} has applied for the first time maximum-entropy methods
to quantum system. The quantum analogue of ME uses the Umegaki relative entropy 
\cite{Umegaki62}. Although Ochs, Ohya and Petz \cite{Ochs75,Ohya93} have given axioms 
for the von Neumann entropy and the Umegaki relative entropy, no axioms of the inference 
are known which would lead to the quantum ME. However, several justifications of the 
quantum ME are known \cite{Balian87,CatichaNC,Streater11} and a few references to 
applications are collected in \cite{Ali12}. 
\par
Let us turn to a more detailed characteristics of the discontinuities in question. 
Given a fixed set of observables and a prior state, the (quantum) ME-inference is, 
in a suitable restriction, a real-analytic map. This restriction has a continuous 
extension for mutually commuting observables, as was shown by Barndorff-Nielsen 
\cite{Barndorff78}, p.\ 154, for the case of probability distributions. So the 
discontinuities are a pure quantum phenomenon, like for example entanglement. We 
\cite{Weis12} have found a pair of non-commutative observables of a three-level 
quantum system, where this continuous extension is not possible and we 
\cite{Weis_continuous} stress that 
\begin{enumerate}
\item
discontinuities are not removable by changing single values of the ME-inference, 
\item
if the true state of the quantum system has full support then the discontinuities 
have no consequences in asymptotic state estimation, except possibly about the 
convergence rate,
\item
the continuity of the inference map is characterized by the openness of the 
restricted linear map assigning expected values to the (fixed) observables,
\item
the openness condition has the physical interpretation of tolerance of an
ME-inference state for small ambiguity of expected values.
\end{enumerate}
In the following we will explain these statements.
%
%
\section{The ME-inference}
\par
We define the ME-inference. We show why discontinuities of the ME-inference
are not removable and we address the asymptotic state estimation.
\begin{figure}
\begin{picture}(12,3)
\put(1,0){\includegraphics[height=3cm, bb=90 110 470 470, clip=]%
{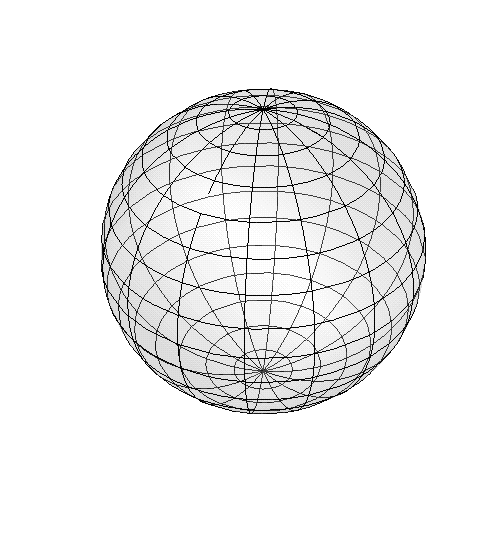}}
\put(5,0){\includegraphics[height=3cm, bb=470 510 1600 1710, clip=]%
{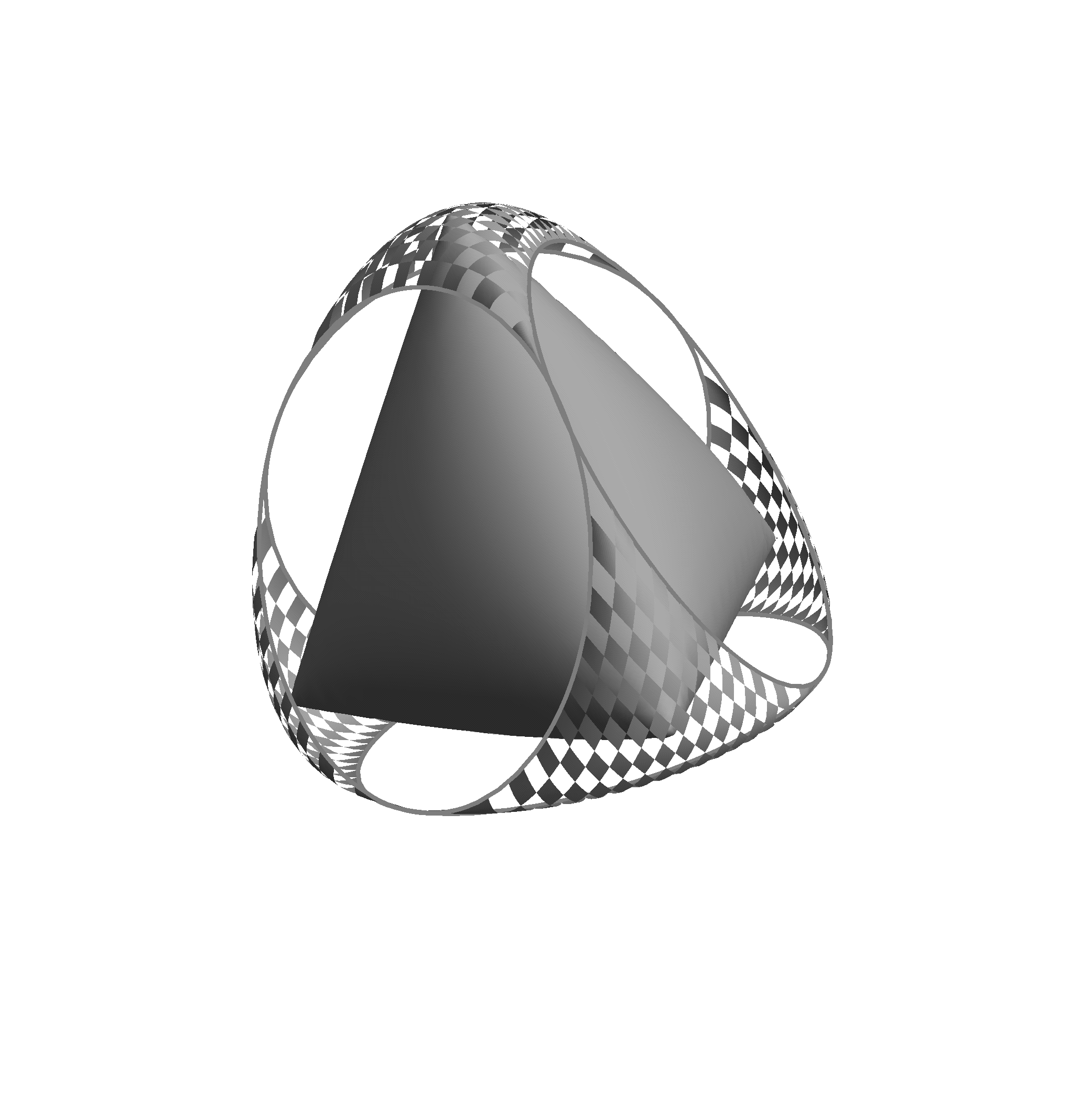}}
\put(9,0){\includegraphics[height=3cm, bb=-15 0 500 530, clip=]%
{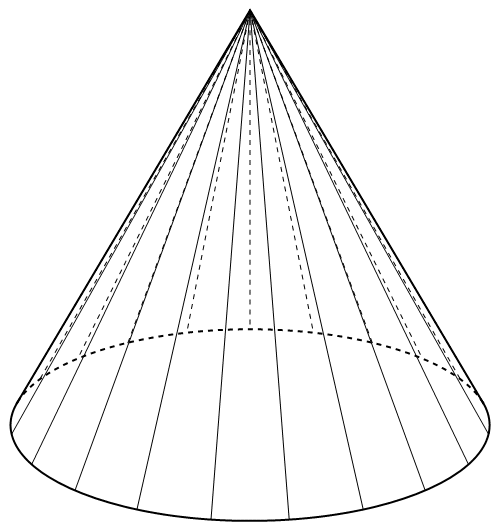}}
\put(0.5,0){a)}
\put(4.5,0){b)}
\put(8.5,0){c)}
\end{picture}
\caption{\label{fig}a) Bloch ball, b) and c) reductions of 
$\cS({\rm Mat}(3,\bC))$.}
\end{figure}
\par
Mathematically, the state of an $n$-level {\it quantum system} is described by 
a linear functional on the C*-algebra ${\rm Mat}(n,\bC)$ of complex 
$n\times n$-matrices, with identity $\id_n$ and zero $0_n$, $n\in\bN$. We will 
use C*-subalgebras $\cA\subset{\rm Mat}(n,\bC)$ with identity $\id_n$, for 
example diagonal matrices $\cA\cong\bC^n$ describe probability distributions on 
the sample space $\{1,\ldots,n\}$. States on $\cA$ are in one-to-one 
correspondence with {\it density matrices} in $\cA$, that is positive 
semi-definite matrices of trace one. The {\it state space} of the algebra $\cA$ 
is the convex body 
\[
\cS=\cS(\cA):=\{\rho\in\cA\mid
\rho\text{ is a density matrix}\}.
\]
We use the terms of density matrix and state synonymously. The state space of a 
two-level quantum system is a three-dimensional Euclidean ball, known as 
{\it Bloch ball}, see Figure~\ref{fig}~a). 
\par
A self-adjoint matrix $a\in\cA$ is also known as {\it observable}, it can be used to 
make measurements on a quantum system. The real vector space 
$\sa:=\{a\in\cA\mid a^*=a\}$ of observables is a Euclidean space with scalar 
product $\langle a,b\rangle:={\rm tr}(ab)$, $a,b\in\sa$. We insist that the identity 
of $\cA$ is $\id_n$. Then an observable $a$ has spectral decomposition 
$a=\sum_{\lambda}\lambda p_\lambda$, where summation runs over all eigenvalues $\lambda$
of $a$ and $p_\lambda=p_\lambda^*=p_\lambda^2$ is an orthogonal projection such that 
$p_\lambda p_{\lambda'}=0$ whenever $\lambda'\neq \lambda$ and
$\sum_\lambda p_\lambda=\id_n$. The eigenvalues of $a$ are 
real. The observable $a$ represents a {\it simple} or 
{\it von Neumann} measurement, where the outcome $\lambda$ is observed with probability 
$\langle\rho,p_\lambda\rangle$, given the quantum system is in the state $\rho\in\cS$. 
Therefore, the {\it expected value} of $a$ is $\langle\rho,a\rangle$, see for example 
\cite{Nielsen}. Given several observables 
\[
a_1,\ldots,a_k, \quad k\in\bN,
\]
we introduce a linear map
\[
\bE:\sa\to\bR^k,\quad
b\mapsto(\langle b,a_1\rangle,\ldots,\langle b,a_k\rangle).
\]
If $\rho\in\cS$ is a density matrix, then we call $\bE(\rho)$ the {\it expected value}
of $\rho$ (the observables are fixed) and we define the convex body
\[
\cC:=\bE(\cS).
\]
\par
The partial information of an expected value $m\in\cC$, may not be sufficient to
specify a quantum state. The {\it (Umegaki) relative entropy} of a state $\rho$ 
from a state $\sigma$ is defined by 
\[
S(\rho,\sigma):=\left\{\begin{array}{ll}
{\rm tr}\,\rho(\log(\rho)-\log(\sigma)) & \text{if } 
{\rm Im}(\rho)\subset{\rm Im}(\sigma),\\
+\infty & \text{else,}
\end{array}\right.
\]
where ${\rm Im}$ is the image of a matrix as a linear map in a fixed basis. A 
self-adjoint matrix $\theta\in\sa$ defines an invertible {\it prior state} 
$\sigma:=e^\theta/{\rm tr}(e^\theta)$. We assume that observables and prior 
are fixed. Then the {\it ME-inference} is defined by
\[
\Psi=\Psi_\theta:
\cC\to\cS,\quad
m\mapsto{\rm argmin}\{S(\rho,\sigma)\mid\bE(\rho)=m, \rho\in\cS\}.
\]
Since $S(\rho,\id_n/n)=-H(\rho)+\log(n)$ holds for $\rho\in\cS$, where
$H(\rho):={\rm tr}(\rho\log(\rho))$ is the {\it von Neumann entropy}, we
call the ME-inference for $\theta=0$ {\it maximum-entropy inference}.
\par
We address the irremovability of discontinuities of $\Psi$. It is well-known 
that all states
\[
R(\lambda)
:=
\frac{\exp(\theta+\lambda_1 a_1+\cdots+\lambda_k a_k)}
{{\rm tr}\,\exp(\theta+\lambda_1 a_1+\cdots+\lambda_k a_k)},\quad
\lambda=(\lambda_1,\ldots,\lambda_k)\in\bR^k
\]
are ME-inference states. The coefficients $\lambda_i$ are Lagrangian multipliers 
for the expected value constraints and $-\lambda_i$ are considered 
{\it generalized inverse temperatures} \cite{Ingarden97}. What is more, if 
$\id_n,a_1,\ldots,a_k$ are linearly independent, then 
\begin{equation}\label{eq:real-analytic}
\bR^k\to\cC^\circ, \quad
\lambda\mapsto\bE\circ R(\lambda)
\end{equation}
is a real-analytic diffeomorphism to the interior $\cC^\circ$ of $\cC$. (If linear 
independence does not hold then the interior has to be replaced by the interior with 
respect to the affine hull of $\cC$.) For the maximum-entropy inference 
(\ref{eq:real-analytic}) was proved by Wichmann \cite{Wichmann63} and we 
\cite{Weis_topology} have proved the real analyticity for arbitrary priors. It is 
known that all elements of the {\it exponential family} $\cE:=R(\bR^k)$ are 
ME-inference states and the inclusion $\Psi(\cC)\subset\overline{\cE}$ of inference 
states into the norm closure $\overline{\cE}$ of the exponential family $\cE$ was
proved \cite{Wichmann63,Weis_continuous}. So
\begin{equation}\label{eq:2inclusions}
\cE=\Psi(\cC^\circ)\subset\Psi(\cC)\subset\overline{\cE}
\end{equation}
follows and (\ref{eq:2inclusions}) shows that a discontinuity of $\Psi$, if present, 
is noticeable by taking the closure of $\cE$. To find the discontinuities of $\Psi$
there is no need to evaluate $\Psi$ on the boundary of $\cC$, where all its
discontinuities lie.
\par
Sample mean values, identified with expected values, can be used to make an
inference from a real experiment. We show that discontinuities of the ME-inference
do not affect the asymptotical inference. Measuring the observable $a$ on $N$ iid 
copies of the state $\rho$ of a quantum system, the {\it sample mean value} of 
$a$ is defined as the arithmetic mean of the $N$ measurement outcomes, $N\in\bN$. 
By the law of large numbers, the sample mean value of $a$ converges to the expected 
value of $a$ for $N\to\infty$. If $kN$ iid copies of $\rho$ are available,
we can measure $N$ times each of the observables $\{a_i\}_{i=1}^k$. Their joint 
sample mean value will converge to $\bE(\rho)\in\bR^k$. Whenever a sample 
mean value lies outside of the set $\cC$ of expected values, we can follow Petz's 
suggestion \cite{Petz08} and apply the (Lipschitz-continuous) method of least squares 
to estimate an expected value $m_N\in\cC$. (More generally, we can use a continuous
estimator which is the identity on $\cC$.)
If $\rho$ has full support, that is $\rho$ is invertible as a linear map, then the 
expected value $\bE(\rho)$ lies in $\cC^\circ$, see for example \cite{Weis_continuous}. 
Therefore, for large $N$, the conversion of a sample mean value into an expected 
value $m_N$ and subsequently into an ME-inference state $\Psi(m_N)$ is, as a 
composition of the continuous estimation, the inverse diffeomorphism 
(\ref{eq:real-analytic}) and the parametrization $R$, a continuous mapping. In 
particular, the ME-inference states $\Psi(m_N)$ converge 
to the state on the exponential family $\cE$ with expected value $\bE(\rho)$ for
$N\to\infty$. It is possible that the convergence rate of the sequence 
$\{\Psi(m_N)\}_{N\in\bN}$ becomes worse the closer $\bE(\rho)$ lies at a 
discontinuity of $\Psi$.
%
%
%
\section{The openness condition}
\par
We explain the openness condition about continuity of the ME-inference, we make 
an example within the simplest possible algebra, and we address a physical 
interpretation of the openness condition.
\par
We call the restricted linear map $\bE|_{\cS}$ {\it open at} $\rho\in\cS$ if for 
all neighborhoods $U\subset\cS$ of $\rho$ the image $\bE(U)$ is a neighborhood of 
$\bE(\rho)$ in $\cC$. Here we use the norm topology, restricted to $\cS$ 
respectively to $\cC$.\\[0.2cm]
{\bf Lemma.} If $\Psi$ is continuous at $m\in\cC$ then
$\bE|_\cS$ is open at $\Psi(m)$.\\[0.2cm]
{\bf Proof.} Let $U$ be a neighborhood of $\Psi(m)$. Then $\Psi^{-1}(U)$ is a 
neigh\-bor\-hood of $m$ by continuity of $\Psi$. By definition of $\Psi$ we 
have $\bE(U)\supset\Psi^{-1}(U)$, completing the proof.\hspace*{\fill}$\square$\\[0.2cm]
The converse of the lemma is true, it follows from the optimality of the 
ME-inference (minimization of the relative entropy) and uniqueness of the minimizer 
\cite{Weis_continuous}. 
\par
We recall the simplest algebra to make an example of a discontinuous ME-inference. 
We have shown \cite{Weis_continuous} that the ME-inference is continuous for all 
balls $\cS$ as well as for all polytopes $\cC$. Since the state space of a two-level 
quantum system is the Bloch-ball and the state space of a commutative algebra is a 
simplex (where $\cC$ is a polytope) a discontinuity of $\Psi$ is only possible for a 
C*-algebra which properly includes the algebra ${\rm Mat}(2,\bC)$ of a two-level 
quantum system. Before we arrive at a suitable three-dimensional state space 
of a subalgebra of $\cA:={\rm Mat}(3,\bC)$, let us recall that the eight-dimensional 
state space $\cS(\cA)$ is not a ball. This is demonstrated by the section of 
$\cS(\cA)$ with the plane $\{\tfrac{\id_3}{3}+\left(\begin{smallmatrix}
0 & x & y\\x & 0 & z\\y & z & 0\end{smallmatrix}\right)\mid
x,y,z\in\bR\}$, depicted\footnote{%
With kind permission of Springer Science+Business Media. Reprinted from 
\cite{Bengtsson11}.}
in Figure~\ref{fig}~b) inside the projection of $\cS(\cA)$ to the same plane. The 
section is an inflated tetrahedron, the projection has four disks on its boundary, 
mutually intersecting in six points \cite{Bengtsson11}. 
\par
The observables we \cite{Weis12} have found generating a discontinuous 
ME-inference, are
\begin{equation}\label{eq:observables-staffelberg}
a_1:=\sigma_1\oplus 0,\qquad
a_2:=\sigma_2\oplus 1,
\end{equation}
interpreted as $3\times 3$-blockdiagonal matrices with block sizes two and one. 
Here we have used Pauli matrices 
$\sigma_1:=\left(\begin{smallmatrix}0 & 1\\1 & 0\end{smallmatrix}\right)$,
$\sigma_2:=\left(\begin{smallmatrix}0 & -\ii\\\ii & 0
\end{smallmatrix}\right)$ and 
$\sigma_3:=\left(\begin{smallmatrix}1 & 0\\0 & -1\end{smallmatrix}\right)$. The 
simplest C*-algebra containing $\id_3,a_1,a_2$ is the direct sum of 
${\rm Mat}(2,\bC)$ and $\bC$ with a state space equal to a four-dimensional cone 
based on the Bloch ball. The matrices $\id_3,a_1,a_2$ belong to a smaller algebra
$\cB$, which is the real *-subalgebra spanned by
\[  
\sigma_1\oplus 0,\qquad
\sigma_2\oplus 0,\qquad
\ii\sigma_3\oplus 0,\qquad
\id_2\oplus 0,\qquad
\id_3.
\]
The algebra $\cB$ is isomorphic to ${\rm Mat}(2,\bR)\oplus\bR$ (by exchanging $\sigma_2$ 
and $\sigma_3$). We extend the above definitions from a C*-algebra to a real *-algebra.
The state space $\cS(\cB)$ is a three-dimensional cone, depicted in Figure~\ref{fig}~c). 
Its directrix (base circle) is parametrized for real $\alpha$ by 
\[
\rho(\alpha):=
\tfrac{1}{2}(\id_2+\sin(\alpha)\sigma_1+\cos(\alpha)\sigma_2)\oplus 0.
\]
Since the exponential family $\cE$, defined previously, is included in $\cS(\cB)$ and 
since $\Psi(\cC)$, computed in the algebra ${\rm Mat}(3,\bC)$, is a subset of the norm 
closure  $\overline{\cE}$ by (\ref{eq:2inclusions}), we have $\Psi(\cC)\subset\cS(\cB)$ 
because $\cS(\cB)$ is norm closed. Therefore the ME-inference with respect to $a_1,a_2$ 
and with prior in $\cB$ can be studied equivalently in the three-dimensional cone 
$\cS(\cB)$ or in the eight-dimensional state space of ${\rm Mat}(3,\bC)$. 
\begin{figure}
\begin{picture}(9,3.5)
\put(1,0){\includegraphics[height=3.5cm, bb=70 40 500 430, clip=]%
{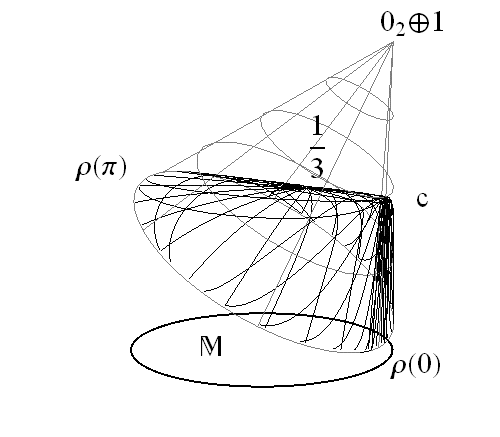}}
\put(5.5,0){\includegraphics[height=3cm, bb=40 20 500 375, clip=]%
{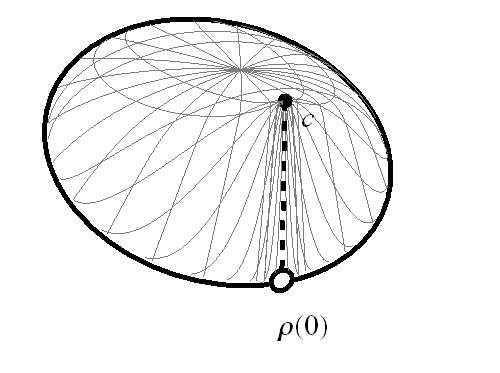}}
\put(0.5,0){a)}
\put(5,0){b)}
\end{picture}
\caption{\label{fig2}The Staffelberg family a) inside the cone
$\cS(\cB)$ and b) extended to $\Psi_0(\cC)$.}
\end{figure}
\par
We compute a full set $\Psi_0(\cC)$ of maximum-entropy inference states.
For the prior $\id_3/3$ and observables $a_1,a_2$ from (\ref{eq:observables-staffelberg}),
the exponential family $\cE$ is called {\it Staffelberg family} in \cite{Weis12}. The 
generatrix (surface line) $[\rho(0),0_2\oplus 1]$ of the cone 
$\cS(\cB)$ is perpendicular to ${\rm span}_\bR\{a_1,a_2\}$, its midpoint is denoted
$c$. See\footnote{Reprinted with permission from \cite{Weis12}.
Copyright 2012, American Institute of Physics.}
Figure~\ref{fig2}~a) for a drawing of $\cE$ inside the cone $\cS(\cB)$ with the set
of expected values underneath, such that $\bE$ acts by vertical projection along  
$[\rho(0),0_2\oplus 1]$. Maximum-entropy inference $\Psi_0(m)$ for $m$ in the interior 
$\cC^\circ$ of expected values covers the exponential family $\cE$. For 
$m_0:=(0,1)=\bE(\rho(0))=\bE(0_2\oplus 1)$ we have $\Psi_0(m_0)=c$ and 
$\Psi_0(m)$ lies on the directrix of $\cS(\cB)$ for boundary points 
$m\neq m_0$. The maximum-entropy inference $\Psi_0:\cC\to\cS$ jumps at $m_0$
\cite{Weis12}. This can be seen\footnotemark[2] in Figure~\ref{fig2}~b), where
$\Psi_0(\cC)$ is depicted, consisting of $\cE$, of the pointed circle 
$\rho(\alpha)$ for $\alpha\in(0,2\pi)$ and of $c$. The segment 
$[\rho(0),c):=\{(1-\lambda)\rho(0)+\lambda c\mid 0\leq\lambda<1\}$, dashed in
the figure, belongs to the norm closure $\overline{\cE}$ and not to $\Psi_0(\cC)$.
\par
\begin{figure}
\begin{picture}(8,3)
\put(1,0){\includegraphics[height=3cm, bb=35 60 500 410, clip=]%
{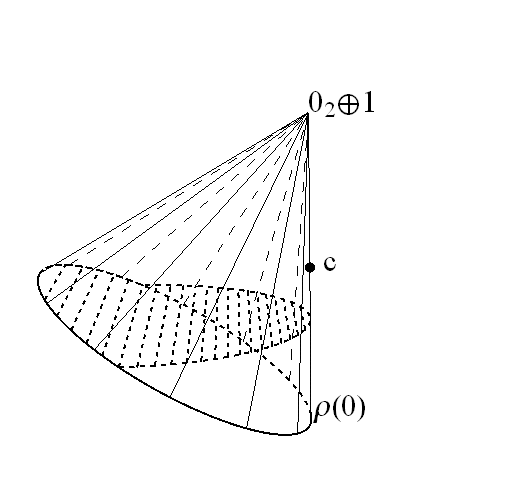}}
\put(5,0){\includegraphics[height=3cm, bb=-50 0 500 500, clip=]%
{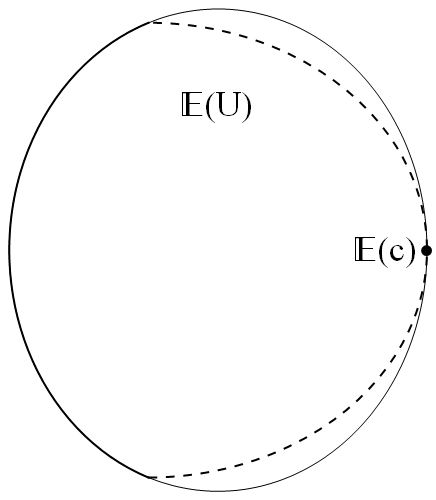}}
\put(0.5,0){a)}
\put(4.5,0){b)}
\end{picture}
\caption{\label{fig3}a) The upper part $U$ above the dashed separation is a 
neighborhood of $c$ in the cone $\cS(\cB)$. b) The image $\bE(U)$ is not a 
neighborhood of $\bE(c)$.}
\end{figure}

We arrive at the same conclusion of a discontinuity at $m_0=(0,1)$ through the 
openness condition. We have $\Psi_0(m_0)=c$ and the neighborhood 
$U(c):=\{\rho\in\cS(\cB)\mid\langle\rho,0_2\oplus 1-\rho(0)\rangle\geq-\tfrac 1{3}\}$, 
shown in Figure~\ref{fig3}~a), proves that $\bE|_{\cS(\cB)}$ is not open at $c$, 
because the image $\bE(U)$ in Figure~\ref{fig3}~b) has, at $\bE(c)=m_0$ a larger
boundary curvature than $\cC$. Hence the above lemma proves that the maximum-entropy 
inference $\Psi_0$ is not continuous at $m_0$.
\par
The openness of $\bE|_\cS$ has a natural interpretation in empirical quantum state 
estimation. Before we can apply the ME-inference to a sample mean value we have to 
estimate an expected value in $\cC$, see the last section. The estimation provides
a finite sequence (indexed by sample size) of expected values  
$m_1,\ldots,m_N\in\cC\subset\bR^k$, $N\in\bN$ and the (theoretical) asymptotical 
sequence converges to the expected value $m:=\bE(\rho)$ of the true state $\rho$ 
of the given quantum system for $N\to\infty$. The openness of $\bE|_\cS$ at 
$\Psi(m)\in\cS$ means that a small ambiguity of $m$, resulting for example from 
the identification of $m=m_N$ for a finite sample length $N\in\bN$, can be balanced 
by a small adjustment of $\Psi(m)$. See \cite{Shirokov10} for this point of view in 
a different context. 
%
%
%
%
%
%

\begin{theacknowledgments}
We would like to thank Arleta Szko\l a and Michael Nussbaum for an interesting 
discussion about maximizing entropies. We would like to thank Adom Giffin and 
Ariel Caticha, as well as the anonymous referee, for valuable correspondences. 
This work is supported by the DFG project ``Quantenstatistik:\ 
Ent\-schei\-dungs\-pro\-bleme und entropi\-sche Funktionale auf Zustandsr\"aumen''.
\end{theacknowledgments}
%


\bibliographystyle{aipproc}   

\end{document}